\documentclass[]{iopart}
\usepackage[latin1]{inputenc}
\usepackage{graphicx}

\newcommand{\avg}[1]{E\,#1}

\begin{document}

\title[Free energy fluctuations and bond chaos]{An exact
relation between free energy fluctuations and bond chaos in the
Sherrington-Kirkpatrick model}
\author{T Aspelmeier}
\address{Max Planck Institute for Dynamics and Self Organization,
37073 Göttingen, Germany}

\begin{abstract}
Using a variant of the interpolating Hamiltonian technique, we show that there
exists, in the Sherrington-Kirkpatrick spin glass, an exact connection
between the sample-to-sample fluctuations of the free energy and bond chaos
involving 2- and 4-replica overlaps between replicas with different
but correlated bonds. This relation is used to derive an upper bound of the
fluctuations.
\end{abstract}




\section{Introduction}

Extreme value statistics is a very active field in current mathematical physics
since the discovery of the Tracy-Widom distribution for the largest
(or smallest) eigenvalue of a Gaussian random matrix, see \cite{Biroli:2007}
for an overview. The Tracy-Widom distribution is believed to consitute a new
universality class for extreme values in addition to the three ``classical''
ones (Weibull, Gumbel, Fr\'echet). There are however many cases which do not
fall in any of these four classes. One important example is the distribution of
ground state energies in the Sherrington-Kirkpatrick model
\cite{Sherrington:1975}. Despite tremendous numerical effort over the years
\cite{Cabasino:1988,Bouchaud:2003,Palassini:2003,Andreanov:2004,Boettcher:2005a,%
Katzgraber:2005,Pal:2006},
there is still no complete agreement as to what kind of distribution the ground
states follow. Analytically, there is no theory (to the best of our knowledge)
which would predict a particular limiting distribution for large system sizes
$N$. Not even the width of the distribution is precisely known: the numerical
simulations seem to suggest that the width scales as $N^{\mu}$ with $\mu\approx
\frac 14$, and this is supported by some heuristic arguments
\cite{Aspelmeier:2003a,Bouchaud:2003}. Other arguments favour $\mu=\frac 16$
\cite{Crisanti:1992,Aspelmeier:2007}. (If this problem fell into the
Tracy-Widom universality class, the width would scale as $N^{1/3}$
\cite{Andreanov:2004}. This seems to be ruled out by the numerical
results.)

In addition to the ground state energies and their sample-to-sample fluctuations
one can also consider the sample-to-sample fluctuations of the free energy at a
finite temperature within the spin glass phase. The natural expectation would be
that in the low temperature phase these free energies fall into the same
universality class as the ground state energies (although this has never been
proved). However, the distribution of the free energies appears as inaccessible
as the one of the ground state energies. Part of the difficulty lies in the fact
that the variance of the distribution scales with a subextensive power of $N$. In
order to calculate subextensive terms, it is usually necessary to go to higher
than the leading order in the loop expansion of the spin glass problem. Due to
the massless modes present throughout the spin glass phase this has so far been
impossible in the Sherrington-Kirkpatrick model. For the finite-dimensional
spin glass, this problem is not so severe and the fluctuations could be
calculated \cite{Wehr:1990,Aspelmeier:2003b}. They are however fundamentally
different from the ones in the Sherrington-Kirkpatrick model, which we will be
considering here.

In this paper we present a way which circumvents this obstacle by constructing an
exact relation between the free energy fluctuations and bond chaos in spin
glasses. This connection has been briefly described in \cite{Aspelmeier:2007a}
and we present the details of the calculation here. Using this relation, the
width of the distribution can in principle be calculated by calculating chaos. A
part of the necessary aspects of chaos has been calculated in
\cite{Aspelmeier:2008bpre}, and the results from that paper will be sufficient to
derive the upper bound $\mu\le\frac 14$ here. For the full answer, it will be
necessary to calculate more complicated objects such as simultaneous $4$-replica
overlaps. We will not be able to solve this formidable problem here.

This paper is organized as follows. In Sec.~\ref{secreview} we briefly review a
few methods and results from the literature in order to compare them with our
own theory later on. We derive the connection to bond chaos in
Sec.~\ref{secinterpolating}. The fluctuations above and at the critical
temperature, as well as the bound $\mu\le\frac 14$ in the low temperature phase
are calculated in Sec.~\ref{secfluctuations}. We end with a conclusion in
Sec.~\ref{secconclusion}.

\section{Above and at the critical temperature}
\label{secreview}
In this section we review a few methods and results above and at the critical
temperature from the literature for completeness and for comparison with our
own results later on.

Analytically, the free energy fluctuations of any disordered system can in 
principle be found with the replica method. Given the partition function $Z$ of 
a system of size $N$, it can easily be shown that a Taylor expansion of $\log 
\overline{Z^n}$ in powers of $n$ yields
\begin{equation}
\log\overline{Z^n} = -n\beta F_N + \frac{n^2}{2}\Delta F_N^2 + \cdots,
\label{replica}
\end{equation}
where the overbar means the average over the disorder, $\beta=1/k_B T$ is the 
inverse temperature, $F_N$ is the average free energy at system size $N$, and 
$\Delta F_N$ denotes its sample-to-sample fluctuations. The dots indicate 
higher order cumulants. Using the replica formalism, one can calculate 
$\overline{Z^n}$ for integer $n$ and try to continue the resulting expression 
to real (or, indeed, complex) $n$ and isolate the coefficient of the second 
order term which represents the fluctuations. In the case of the Ising spin 
glass this works very nicely above and at the critical temperature. It is 
straightforward to show with the standard replica formalism for the mean-field 
spin glass \cite{Mezard:1987} that in the high temperature phase ($\beta<1$), 
where the saddle point is replica symmetric and its Hessian has only strictly 
positive eigenvalues, the fluctuations are 
\begin{equation}
\beta^2\Delta F_N^2 = -\frac 12 
\log(1-\beta^2)-\frac{\beta^2}{2}+\mathcal O(1/N)
\label{fluctabove}
\end{equation} 
\cite{Crisanti:1992,Parisi:1993}. As the critical temperature $T_c$ is
approached ($\beta\nearrow 1/T_c=1$), this expression diverges, which indicates that the 
fluctuations at the critical point must also diverge with $N$. A 
straightforward extension of the calculation in \cite{Parisi:1993} shows that 
the fluctuations at the critical point are 
\begin{equation}
\beta^2\Delta F_N^2 = \frac 16 \log N 
+ \mathcal O(1),
\label{fluctat}
\end{equation}
 which does indeed diverge as $N\to\infty$. As a check, we can rederive this
 result from Eq.~(\ref{fluctabove}) by
 isolating the divergent part, $-\frac 12 \log(1-\beta)$, and replacing $1-\beta\sim
 \tau$ (where $\tau=\frac{T-T_c}{T_c}$ is the reduced temperature) by
 $xN^{-1/3}$. The variable $x=\tau N^{1/3}$ is the correct scaling combination
 in the critical region \cite{Parisi:1993,Yeo:2005}. Keeping $x$ fixed and
 letting $N$ tend to infinity in $-\frac 12 \log(xN^{-1/3})$ results in
 Eq.~(\ref{fluctat}).

In the low temperature phase, the situation is much more complex and there are 
no reliable analytical results.

\section{Interpolating Hamiltonian}
\label{secinterpolating}

In this section we will derive two different exact expressions for the
fluctuations in terms of chaos using interpolating Hamiltonians. While the
calculation presented here is in spirit similar to the one by Billoire
\cite{Billoire:2006}, there is an important difference. Here, we do not
interpolate between a big system and two small systems (see also
\cite{Guerra:2002}) but between two equally big systems. This may seem strange at
first sight but is in fact the key to making any analytical progress on this
particular problem.

\subsection{First route to chaos}
Consider the following interpolating Hamiltonian:
\begin{equation}
  \mathcal H_t = -\sqrt{\frac{1-t}{N}}\sum_{i<j}J_{ij}s_i s_j - 
  \sqrt{\frac{t}{N}}\sum_{i<j}J'_{ij}s_i s_j 
\label{hamiltonian1}
\end{equation}
with $N$ Ising spins $s_i$, $0\le t\le 1$ and $J_{ij}$, $J'_{ij}$ independent
Gaussian random variables with unit variance. The parameter $t$ interpolates
between one spin glass system ($t=0$) and a statistically independent, but
otherwise identical one at $t=1$. It is important to note that also for each
other value of $t$ the Hamiltonian describes a normal spin glass, the coupling
constants being $\sqrt{1-t}J_{ij}+\sqrt{t}J'_{ij}$ which are Gaussian random
variables with unit variance.

The partition function of this Hamiltonian is $Z_t=\Tr\exp(-\beta \mathcal H_t)$
and the free energy is $\beta F_t=-\log Z_t$. The sample-to-sample fluctuations
of the free energy of the SK model can be obtained in the following way. Denoting
the average over all coupling constants $J_{ij}$ and $J'_{ij}$ (and later also
$J''_{ij}$ and others) by $\avg{\cdots}$, we have
\begin{eqnarray}
\avg{(\log Z_1-\log Z_0)^2} &= \beta^2 \avg{(F_1-F_0)^2}\\
 &= \beta^2(\avg{F_1^2} - 2\avg{F_1 F_0}  + \avg{F_0^2})\\
 &= 2\beta^2(\overline{F^2}-\overline{F}^2)\\
 &= 2\beta^2\Delta F_N^2.
 \label{interpolation1}
\end{eqnarray}
The penultimate step follows from the fact that $\avg{F_1^2} = \avg{F_0^2}
=:\overline{F^2} $ is the disorder average of the squared spin glass free energy
and that the average $\avg{F_1 F_0}=(\avg{F_1}) (\avg{F_0})=:\overline{F}^2$
factorizes into the square of the averaged free energy since the coupling constants in the
two Hamiltonians $\mathcal H_0$ and $\mathcal H_1$ are independent. Using this 
formulation and the idea developed in \cite{Guerra:2002} to
represent $\log Z_1-\log Z_0$ by differentiating with respect to the
interpolation parameter $t$ and immediately integrating again, the fluctuations
can be written as
\begin{equation}
  2\beta^2\Delta F_N^2  = \avg{(\log Z_1-\log
  Z_0)^2} = \int_0^1 dt\int_0^1
  d\tau\,\avg{\frac{\partial}{\partial t}\log
  Z_t\frac{\partial}{\partial \tau}\log Z_\tau}.
  \label{route1}
\end{equation}

In \ref{approute1} it is shown how to manipulate this expression in
order to arrive at Eq.~(\ref{average4}), which is repeated here for convenience.
Note that this equation is exact.
\begin{eqnarray}
\fl
\avg{\frac{\partial}{\partial t}\log Z_t\frac{\partial}{\partial
      \tau}\log Z_\tau} = \frac{N^2\beta^4}{16}     
      \left(2-\frac{\sqrt{1-t}\sqrt{\tau}}{\sqrt{t}\sqrt{1-\tau}}
             -\frac{\sqrt{1-\tau}\sqrt{t}}{\sqrt{\tau}\sqrt{1-t}}\right)
        \avg{\langle(q_{13}^2-q_{14}^2)(q_{13}^2-q_{23}^2)\rangle}\nonumber\\
      + \frac{N \beta^2}{4\sqrt{t\tau}} \left(\avg{\langle q_{13}^2\rangle} -
      \frac 1N\right).
     \label{average4a}
\end{eqnarray}
The symbols $q_{ab}$ are overlaps between independent replicas with different
interpolation parameters,
\begin{equation}
q_{ab}(t,\tau) = \frac 1N \sum_i s_i^{a,t} s_i^{b,\tau}.
\end{equation}
In Eq.~(\ref{average4a}) replicas $1$ and $2$ have parameter $t$ and replicas
$3$ and $4$ have parameter $\tau$. The angular brackets $\langle\cdots\rangle$
denote the thermal average of a system of independent replicas with the
appropriate interpolation parameters.

We have thus established a connection between the fluctuations and the overlap
between replicas with different interpolation parameters. The last important step
is to realize that for any given value of $t$, $\mathcal H_t$ represents a normal
mean-field spin glass with Gaussian couplings just like any other. The overlap
$q_{13}$ between two replicas with different interpolation parameters is
therefore an overlap between two normal spin glasses with identical bonds (if
$t=\tau$), uncorrelated bonds (if $t=0, \tau=1$ or vice versa) or related, but
not equal bonds (for anything in between). This immediately shows the connection to
chaos in spin glasses. Chaos concerns the question how the equilibrium states of
two initially equal systems are related when a small perturbation is applied to
one of them, e.g.\ a small change of temperature (temperature chaos) or a perturbation
of the bonds (bond chaos). When there is chaos, the equilibrium states are
completely unrelated and the overlap is $0$ (in the thermodynamic limit), no
matter how small the perturbation. In our case, we are dealing with bond chaos.

Let $t$ and $\tau$ be given and let the coupling constants of
$\mathcal H_t$ be the reference configuration of bonds:
$K_{ij}^0:=\sqrt{1-t}J_{ij}+\sqrt{t}J'_{ij}$. The coupling constants belonging
to $\mathcal H_\tau$ are $K_{ij}=\sqrt{1-\tau}J_{ij}+\sqrt{\tau}J'_{ij}$.
Since the $J_{ij}$ and $J'_{ij}$ are Gaussian random variables, so are $K^0_{ij}$ and
$K_{ij}$ (also with unit variance). Their
correlation is $E_{J,J'}{K^0_{ij}K_{ij}}=
\sqrt{1-t}\sqrt{1-\tau}+\sqrt{t\tau}$. Instead of using $J_{ij}$ and $J'_{ij}$
as the basic independent random variables one could also use $K^0_{ij}$,
introduce new Gaussian random variables $K'_{ij}$ and take $K^0_{ij}$ and
$K'_{ij}$ as the building blocks of the random variables. We can then write the
bonds pertaining to $\mathcal H_\tau$ as
\begin{equation}
  K_{ij}(\epsilon) = \frac{K^0_{ij}}{\sqrt{1+\epsilon^2}} +\frac{\epsilon
    K'_{ij}}{\sqrt{1+\epsilon^2}},
\end{equation}
such that the correlation between $K^0_{ij}$ and $K_{ij}(\epsilon)$ is
$\avg{K^0_{ij}K_{ij}(\epsilon)}=\frac{1}{\sqrt{1+\epsilon^2}}$. In order
that the bonds $K_{ij}(\epsilon)$ are statistically equivalent to the original
bonds of $\mathcal H_\tau$, the correlation must be equal to the correlation
obtained before, so
\begin{equation}
\frac{1}{\sqrt{1+\epsilon^2}}=\sqrt{1-t}\sqrt{1-\tau}+\sqrt{t\tau}.
\label{epsilondef1}
\end{equation}
Thus we see that the disorder average of the overlap $q_{13}(t,\tau)$ is only a
function of the ``distance'' $\epsilon$ of the coupling constants, i.e.\
$\avg{\langle q_{13}^2(t,\tau)\rangle} =
\avg{\langle q_{13}^2(\epsilon)\rangle}$ 
is only a function of $\epsilon$, not of $t$ and $\tau$ indepently. The same
applies of course for products of overlaps such as
$\avg{\langle q_{13}^2(t,\tau)q_{23}^2(t,\tau)\rangle} =
\avg{\langle q_{13}^2(\epsilon)q_{23}^2(\epsilon)\rangle}$. The distance
$\epsilon$ varies between $0$ and $\infty$.

In order to obtain the fluctuations, we must integrate Eq.~(\ref{average4b})
over $t$ and $\tau$, according to Eq.~(\ref{route1}). But since the overlaps
only depend on $\epsilon$, it is useful to make a variable substitution and go
over to $\epsilon$ and $z:=\sqrt{1+\epsilon^2}\sqrt{\tau}$. We first note that
the integral $\int_0^1 dt\int_0^1 d\tau\,\bullet$ can be restricted to the
range $\tau\le t$ due to symmetry, provided a factor of $2$ is inserted. We can
then make the substitution and obtain
\begin{equation}
\fl
\beta^2\Delta F_N^2 = -\frac{N^2\beta^4}{16}\int_0^\infty d\epsilon \,
f_1(\epsilon)
\avg{\langle(q_{13}^2-q_{14}^2)(q_{13}^2-q_{23}^2)\rangle} +
\frac{N \beta^2}{4}\int_0^\infty d\epsilon\,g_1(\epsilon)\left(\avg{\langle
q_{13}^2\rangle} - \frac 1N\right)
     \label{fluct1}
\end{equation}
where
\begin{eqnarray}
f_1(\epsilon) &= \int_0^1
dz\,\mathcal J\times
\left(\frac{\sqrt{1-t}\sqrt{\tau}}{\sqrt{t}\sqrt{1-\tau}}
             +\frac{\sqrt{1-\tau}\sqrt{t}}{\sqrt{\tau}\sqrt{1-t}}-2\right),
\label{funcf1a}\\
g_1(\epsilon) &= \int_0^1
dz\,\mathcal J \times\frac{1}{\sqrt{t\tau}}
\label{funcg1a}
\end{eqnarray}
with the Jacobian
\begin{equation}
\mathcal J = \frac{4z}{(1+\epsilon^2)^4}
(\epsilon\sqrt{1+\epsilon^2-z^2}+z)(\sqrt{1+\epsilon^2-z^2}-\epsilon z).
\end{equation}
The old variables $t$ and $\tau$, expressed in terms of the new
ones, are
\begin{eqnarray}
t &= \left(\frac{\epsilon\sqrt{1+\epsilon^2-z^2}+z}{1+\epsilon^2}\right)^2, \\
\tau &= \frac{z^2}{1+\epsilon^2}.
\end{eqnarray}
The integrals in Eqs.~(\ref{funcf1a}) and (\ref{funcg1a}) can be evaluated
explicitly and we find
\begin{eqnarray}
f_1(\epsilon) &= 
\frac{4\epsilon^2}{(1+\epsilon^2)^2}\arcsin\frac{1}{\sqrt{1+\epsilon^2}} 
\label{funcf1b}\\
g_1(\epsilon) &=
\frac{2}{(1+\epsilon^2)^{3/2}}\arcsin\frac{1}{\sqrt{1+\epsilon^2}}.
\label{funcg1b}
\end{eqnarray}

Eq.~(\ref{fluct1}) is our first important result. It is exact and connects the
fluctuations with bond chaos. If it were possible to calculate bond chaos (and it
was shown in \cite{Aspelmeier:2008bpre} that at least for the $2$-replica
overlaps it is possible), the flucutations follow immediately since the functions
$f_1(\epsilon)$ and $g_1(\epsilon)$ are ``harmless'' (Eqs.~(\ref{funcf1b}) and
(\ref{funcg1b})). Note that $f_1(\epsilon)$ and $g_1(\epsilon)$ are nonnegative
and $\langle(q_{13}^2-q_{14}^2)(q_{13}^2-q_{23}^2)\rangle$ is also nonnegative (this
is shown in \ref{approute1}). The first term in
Eq.~(\ref{fluct1}) is therefore negative. Hence the second term is an upper bound
for the fluctuations.

\subsection{Second route to chaos}

There is another way to represent the fluctuations with
interpolating Hamiltonians than Eq.~(\ref{interpolation1}) which will lead to a
second expression for the fluctuations. Introducing the Hamilontian $\mathcal
H'_t$ defined by
\begin{equation}
  \mathcal H'_t = -\sqrt{\frac{1-t}{N}}\sum_{i<j}J_{ij}s_i s_j - 
  \sqrt{\frac{t}{N}}\sum_{i<j}J''_{ij}s_i s_j 
\label{hamiltonian2}
\end{equation}
which only differs from $\mathcal H_t$ by the second set of coupling constants
$J''_{ij}$ which are again independent Gaussian random variables with unit
variance, we can write
\begin{eqnarray}
\fl
\avg{(\log Z_1-\log Z_0)(\log Z'_1-\log Z'_0)} &=
\beta^2\avg{(F_1-F_0)(F'_1-F'_0)}\\ 
&= \beta^2 \avg{ (F_1 F'_1 -
 F_1 F'_0 -  F_0 F'_1 +  F_0 F'_0)}\\ 
 &=
\beta^2(\overline{F^2}-\overline{F}^2)\\ &=\beta^2\Delta F_N^2,
 \label{interpolation2}
\end{eqnarray}
where $Z'_t$ and $F'_t$ are the partition function and the free energy
pertaining to $\mathcal H'_t$. The fluctuations can be represented by a double
integral, as above,
\begin{equation}
\beta^2\Delta F_N^2 = \int_0^1 dt\int_0^1
  d\tau\,\avg{\frac{\partial}{\partial t}\log
  Z_t\frac{\partial}{\partial \tau}\log Z'_\tau}.
  \label{route2}
\end{equation}
Proceeding precisely as above and in Appendix \ref{approute1}, we get 
\begin{eqnarray}
\fl
\avg{\frac{\partial}{\partial t}\log Z_t\frac{\partial}{\partial
      \tau}\log Z'_\tau} = \frac{N^2\beta^4}{16}
        \avg{\langle(q_{13}^2-q_{14}^2)(q_{13}^2-q_{23}^2)\rangle}\nonumber\\
      + \frac{N \beta^2}{8\sqrt{1-t}\sqrt{1-\tau}} \left(\avg{\langle
      q_{13}^2\rangle} - \frac 1N\right).
     \label{average4b}
\end{eqnarray}
Replicas $1$ and $2$ have Hamiltonian $\mathcal H_t$ and replicas $3$ and $4$
have $\mathcal H'_\tau$.

Integrating over $t$ and $\tau$ gives us the fluctuations, and again the
overlaps do not depend on $t$ and $\tau$ separately but only on the distance
$\epsilon$. The distance is here not given by Eq.~(\ref{epsilondef1}) but is
slightly different due to the independence of the $J'$s and $J''$s. 
Arguing similarly as above, $\epsilon$ is found to be
related to $t$ and $\tau$ by
\begin{equation}
\frac{1}{\sqrt{1+\epsilon^2}} = \sqrt{1-t}\sqrt{1-\tau}.
\label{epsilondef2}
\end{equation}
Making a change of variables to eliminate $\tau$ in favour of $\epsilon$ yields
\begin{eqnarray}
\fl
\beta^2 \Delta F_N^2 &= \frac{N^2\beta^4}{16}\int_0^\infty d\epsilon
\int_0^{\epsilon^2/(1+\epsilon^2)} dt\,
     \frac{2\epsilon}{(1-t)(1+\epsilon^2)^2} 
     E
     \langle(q_{13}^2-q_{14}^2)(q_{13}^2-q_{23}^2)\rangle\nonumber\\
     \fl
     &\quad + \frac{N \beta^2}{8}\int_0^\infty d\epsilon
     \int_0^{\epsilon^2/(1+\epsilon^2)}dt 
     \frac{2\epsilon}{(1-t)(1+\epsilon^2)^2} 
     \sqrt{1+\epsilon^2}
     \left(E
     \langle q_{13}^2\rangle - \frac 1N\right),
\end{eqnarray}
such that
\begin{equation}
     \fl
\beta^2 \Delta F_N^2 = \frac{N^2\beta^4}{16}\int_0^\infty d\epsilon \,
     f_2(\epsilon) E
     \langle(q_{13}^2-q_{14}^2)(q_{13}^2-q_{23}^2)\rangle
     + \frac{N \beta^2}{4}\int_0^\infty d\epsilon\,
     g_2(\epsilon) \left(E
     \langle q_{13}^2\rangle - \frac 1N\right).
     \label{fluct2}
\end{equation}
This is the second result for the fluctuations. It has precisely the same
structure as Eq.~(\ref{fluct1}). The only difference are the weight functions
under the integrals, which are given by
\begin{eqnarray}
f_2(\epsilon) &= \frac{2\epsilon\log(1+\epsilon^2)}{(1+\epsilon^2)^2} \\
g_2(\epsilon) &= \frac{\epsilon\log(1+\epsilon^2)}{(1+\epsilon^2)^{3/2}},
\end{eqnarray}
and the sign of the first term, which here is positive.

\section{Calculation of the fluctuations}
\label{secfluctuations}

Having established the connection to chaos, we can proceed to calculate the
fluctuations by calculating $\avg{\langle q_{13}^2 \rangle}$ and
$\avg{\langle(q_{13}^2-q_{14}^2)(q_{13}^2-q_{23}^2)\rangle}$. The former can be
accomplished by taking the bond averaged probability distribution $P_\epsilon(q)$
of the overlap $q$ for bond chaos, which has been calculated in
\cite{Aspelmeier:2008bpre}. Averages taken with this probability distribution
will be denoted by $[\cdots]_0$. The latter is more difficult and will be postponed to
a later publication. It requires the joint probability distributions
$P_\epsilon^{123}(q_{13}, q_{23})$ and $P_\epsilon^{1234}(q_{13}, q_{24})$.
However, above and at the critical temperature, there is no replica symmetry
breaking, hence these probability distributions factorize into
$P_\epsilon^{123}(q_{13},q_{23})=P_\epsilon(q_{13})P_\epsilon(q_{23})$ and
$P_\epsilon^{1234}(q_{14},q_{23})=P_\epsilon(q_{14})P_\epsilon(q_{23})$ such
that
\begin{equation}
\avg{\langle(q_{13}^2-q_{14}^2)(q_{13}^2-q_{23}^2)\rangle} =
[q^4]_0 - [q^2]_0^2.
\end{equation}
We will be able to calculate this. Below the critical temperature, on the other
hand, we will have to content ourselves with an upper bound of the fluctuations
which is given by the second integral in Eq.~(\ref{fluct1}).

\subsection{Above the critical temperature}

The nonnormalised probability distribution of the overlap $q$ for bond chaos,
$R_\epsilon(q)$, above the critical temperature is \cite{Aspelmeier:2008bpre}
\begin{equation}
R_\epsilon^0(q) = e^{-N(\frac{q^2}{2}h(\epsilon)+\mathcal O(q^4))}
\end{equation}
with $h(\epsilon)=1-\frac{\beta^2}{\sqrt{1+\epsilon^2}}$. 

For large $N$, we can easily calculate $[q^2]_0$
and $[q^4]_0$ via steepest descents. At leading
order, the terms of order $q^4$ and higher in the exponent do not contribute. 
Defining $q_n:= \int_0^\infty dq\,q^n R_\epsilon(q)$ (the upper bound may be
set to $\infty$ as this only introduces exponentially small errors), we get
\begin{equation}
q_n = \frac 12 \left(\frac N2
h(\epsilon)\right)^{-(n+1)/2}
\Gamma\left(\frac{n+1}{2}\right)
\end{equation}
such that
\begin{eqnarray}
[q^2]_0 = \frac{q_2}{q_0} = \frac{1}{N h(\epsilon)}, \\ 
{}[q^4]_0 = \frac{q_4}{q_0} = \frac{3}{N^2 h^2(\epsilon)}.
\end{eqnarray}
This allows us to write down two equations for the fluctuations from
our two routes to chaos, Eqs.~(\ref{fluct1}) and (\ref{fluct2}), namely
\begin{eqnarray}
\beta^2\Delta F_N^2 &=-\frac{\beta^4}{8}\int_0^\infty d\epsilon \,
     \frac{f_1(\epsilon)}{h^2(\epsilon)}
     + \frac{\beta^2}{4}\int_0^\infty d\epsilon\,
     g_1(\epsilon) \left(\frac{1}{h(\epsilon)} - 1\right) 
     \label{fluctabove2a}\\
     &= \frac{\beta^4}{8}\int_0^\infty d\epsilon \,
     \frac{f_2(\epsilon)}{h^2(\epsilon)}
     + \frac{\beta^2}{4}\int_0^\infty d\epsilon\,
     g_2(\epsilon) \left(\frac{1}{h(\epsilon)} - 1\right).
     \label{fluctabove2b}
\end{eqnarray}
The second of these expressions can be evaluated explictly, with the result
\begin{equation}
\beta^2 \Delta F_N^2 = -\frac 12 \log(1-\beta^2)-\frac{\beta^2}{2},
\end{equation}
in accordance with Eq.~(\ref{fluctabove}). The author is currently unable to
calculate the integrals in Eq.~(\ref{fluctabove2a}) but numerical checks show
that they give precisely the same result.

\subsection{At the critical temperature}

The nonnormalised probability distribution $R_\epsilon(q)$ precisely at the
critical temperature is given by \cite{Aspelmeier:2008bpre}
\begin{equation}
R_\epsilon(q) = \left\{
\begin{array}{l@{\hspace{1cm}}l}
e^{-Nwq^3/6} & \epsilon\ll N^{-1/6} \\
e^{-Nq^2 h(\epsilon)/2}& N^{-1/6}\ll\epsilon
\end{array}\right. .
\end{equation}
Define as before $q_n=\int_0^\infty dq\, q^n R_\epsilon(q)$. Then we find
\begin{equation}
q_n = \left\{
\begin{array}{l@{\hspace{1cm}}l}
\frac 13 \left(\frac{Nw}{6}\right)^{-(n+1)/3}\Gamma(\frac{n+1}{3}) & \epsilon\ll
N^{-1/6} \\
\frac 12 \left(\frac{N}{2}h(\epsilon)\right)^{-(n+1)/2}\Gamma(\frac{n+1}{2}) &
N^{-1/6}\ll\epsilon
\end{array}
\right.,
\end{equation}
such that 
\begin{equation}
[q^2]_0 = \frac{q_2}{q_0} = \left\{
\begin{array}{l@{\hspace{1cm}}l}
 \left(\frac{Nw}{6}\right)^{-2/3}\frac{1}{\Gamma(\frac 13)} & \epsilon\ll
 N^{-1/6} \\ 
 \left(\frac{N}{2}h(\epsilon)\right)^{-1}\frac{\Gamma(\frac
 32)}{\Gamma(\frac 12)} & N^{-1/6}\ll\epsilon
\end{array}
\right.
\end{equation}
and
\begin{equation}
[q^4]_0 = \frac{q_4}{q_0} = \left\{
\begin{array}{l@{\hspace{1cm}}l}
 \left(\frac{Nw}{6}\right)^{-4/3}\frac{\Gamma(\frac 53)}{\Gamma(\frac 13)} &
 \epsilon\ll N^{-1/6} \\ 
 \left(\frac{N}{2}h(\epsilon)\right)^{-2}\frac{\Gamma(\frac 52)}{\Gamma(\frac
 12)} & N^{-1/6}\ll\epsilon
\end{array}
\right. .
\end{equation}
Now we can evaluate Eq.~(\ref{fluct2}). Plugging $[q^2]_0$ into the second term
of that equation yields a constant of order $1$ which is not of interest and
will therefore not be calculated explicitly. The first term, however, is important. Splitting
the integral into two parts we get asymptotically
\begin{eqnarray}
\fl
\frac{N^2\beta^4}{16}\int_0^\infty d\epsilon\,f_2(\epsilon)([q^4]_0-[q^2]_0^2)
= \frac{N^2\beta^4}{16}\int_0^{N^{-1/6}}d\epsilon\, f_2(\epsilon) 
\frac{\Gamma(\frac 53)\Gamma(\frac
13)-1}{\Gamma^2(\frac 13)}
\left(\frac{Nw}{6}\right)^{-4/3} \\ +
\frac{N^2\beta^4}{16}\int_{N^{-1/6}}^{\infty}d\epsilon\,
f_2(\epsilon) \frac 12 \left(\frac{N}{2}h(\epsilon)\right)^{-2}.
\end{eqnarray}
The first of these integrals yield a constant (independent of $N$). The second one, however, gives a
logarithm at the lower bound such that we get (with $f_2(\epsilon) =
2\epsilon^3 + \mathcal O(\epsilon^5)$,  $h(\epsilon)=\epsilon^2/2 + \mathcal
O(\epsilon^4)$ and $\beta=1$ as we are at the critical point)
\begin{equation}
\beta^2\Delta F_N^2 = \frac 16 \log N + \mathcal O(1).
\end{equation}
This is precisely the known result.

It is interesting to note that we would not have been able to obtain this
result so easily from our first route to chaos, Eq.~(\ref{fluct1}), as both
integrals in that expression grow with some power of $N$, and only their difference cancels out
the leading behaviour and leaves a logarithmic divergence. In order to
actually calculate this, we would need subleading corrections to the integrals,
which would be very hard to obtain indeed.

\subsection{Below the critical temperature}

Now we turn to the low temperature phase. We will not be able
here to solve the complete problem since $P_\epsilon^{123}(q_{13},q_{23})$ and
$P_\epsilon^{1234}(q_{14},q_{23})$ do not factorize in the symmetry breaking
phase. In \cite{Parisi:2000,Guerra:1996} it has been shown how to break down
these probability distributions but the results only apply for $\epsilon=0$.
Instead, we focus on the second term in Eq.~(\ref{fluct1}) since it only requires
$P_\epsilon(q_{13})$ and provides an upper bound for the fluctuations.

From \cite{Aspelmeier:2008bpre} we get the nonnormalised probability
distribution of $q$ in the low temperature phase, which is
\begin{equation}
R_\epsilon(q) = \left\{
\begin{array}{l@{\hspace{1cm}}l}
\hat\theta(q-q_{\mathrm{EA}}) & \epsilon\ll N^{-1/2} \\
e^{-Nc_1 \epsilon^2 q^3} & N^{-1/2}\ll\epsilon\ll N^{-1/5} \\
e^{-Nc_2 \epsilon^3 q^2} & N^{-1/5}\ll\epsilon\le\epsilon_0 \\
e^{-Nf(\epsilon) q^2} & \epsilon_0<\epsilon
\end{array}
\right. ,
\end{equation}
where
\begin{equation}
\hat\theta(x) =
\left\{
\begin{array}{l@{\hspace{1cm}}l}
1 & x<0 \\
e^{-Nc_0x^3} & x>0
\end{array}
\right.
\end{equation} 
with some (unimportant) positive constant $c_0$ and $q_{\mathrm{EA}}$ is the
Edwards-Anderson order parameter, such that
\begin{eqnarray}
q_n &= \left\{
\begin{array}{l@{\hspace{1cm}}l}
\frac{q_{\mathrm{EA}}^{n+1}}{n+1} & \epsilon\ll N^{-1/2} \\
\frac 13 (Nc_1\epsilon^2)^{-(n+1)/3}\Gamma\left(\frac{n+1}{3}\right) &
N^{-1/2}\ll\epsilon\ll N^{-1/5} \\ 
\frac 12 (Nc_2\epsilon^3)^{-(n+1)/2}\Gamma\left(\frac{n+1}{2}\right) &
N^{-1/5}\ll\epsilon\le \epsilon_0 \\ 
\frac 12 (Nf(\epsilon))^{-(n+1)/2}\Gamma\left(\frac{n+1}{2}\right) &
 \epsilon_0<\epsilon
\end{array}
\right. 
\end{eqnarray}
and
\begin{eqnarray}
[q^2]_0 &\propto \left\{
\begin{array}{l@{\hspace{1cm}}l}
\mathrm{const.} & \epsilon\ll N^{-1/2} \\
(N\epsilon^2)^{-2/3} & N^{-1/2}\ll\epsilon\ll N^{-1/5} \\ 
(N\epsilon^3)^{-1} & N^{-1/5}\ll\epsilon\le \epsilon_0 \\ 
(Nf(\epsilon))^{-1} & \epsilon_0<\epsilon
\end{array}
\right. .
\end{eqnarray}
Note the discussion in \cite{Aspelmeier:2008bpre} about why the probability
distribution for $\epsilon\ll N^{-1/2}$ does not coincide with the true
distribution for the Sherrington-Kirkpatrick model. However, this
discrepancy only changes the \textit{value} of $[q^2]_0$ for small $\epsilon$.
It does not change the qualitative behaviour of $[q^2]_0$ as a function of
$\epsilon$.

We can estimate the integral $\frac{N \beta^2}{4}\int_0^\infty
d\epsilon\,g_1(\epsilon)\left(E \langle q_{13}^2\rangle - \frac 1N\right)$ from
Eq.~(\ref{fluct1}) by first neglecting the $1/N$-term under the integral as we
are only interested in the leading behaviour. We can also neglect the
contribution of the integration from $\epsilon_0$ to $\infty$ since it will only
be of order $1$. We also note that we can combine the regions $\epsilon\ll
N^{-1/2}$ and $N^{-1/2}\ll \epsilon\ll N^{-1/5}$ by writing $[q^2]_0 = \mathcal
F(N^{1/2}\epsilon)$ with a scaling function $\mathcal F(x)$ with the properties
$\mathcal F(x)\to\mathrm{const.}$ ($x\to 0$) and $\mathcal F(x)\sim x^{-4/3}$
($x\to\infty$). We then obtain for the first part of the integral (expanding the
function $g_1(\epsilon)$ for small $\epsilon$)
\begin{eqnarray}
\frac{N \beta^2}{4}\int_0^{N^{-1/5}}
d\epsilon\,g_1(\epsilon) [q^2]_0 \approx \frac{N \beta^2}{4}
\pi \int_0^{N^{-1/5}} d\epsilon\, \mathcal F(N^{1/2}\epsilon) \nonumber\\
= N^{1/2}\frac{\beta^2}{4}\pi\int_0^{N^{3/10}} dx\,\mathcal F(x)
\sim N^{1/2}.
\end{eqnarray}
The next part of the integral is
\begin{eqnarray}
\frac{N \beta^2}{4}\int_{N^{-1/5}}^{\epsilon_0}
d\epsilon\,g_1(\epsilon) [q^2]_0 &\sim \frac{N \beta^2}{4}\pi 
\int_{N^{-1/5}}^{\epsilon_0} \frac{1}{N\epsilon^3} \sim N^{2/5}.
\end{eqnarray}
This contribution is smaller than the one we just had and may be
neglected.

The final answer for the fluctuations in the low temperature phase is therefore
\begin{equation}
\beta^2\Delta F_N^2 \le \mathrm{const.}\times N^{1/2},
\end{equation}
i.e.\ we get the upper bound
\begin{equation}
\mu\le\frac 14.
\end{equation}

\section{Conclusion}
\label{secconclusion}

We have shown that the free energy fluctuations in the Sherrington-Kirkpatrick
model can be expressed in two different ways in terms of bond chaos,
Eqs.~(\ref{fluct1}) and (\ref{fluct2}), both of which are exact. The first
formulation consists of a difference of two positive terms while the second is
a sum of positive terms. We have derived an upper bound of the fluctuations
using the first formulation, resulting in $\mu\le \frac 14$. In the future, the
second formulation will be more useful because it allows direct access to the
fluctuations when $4$-replica overlaps are calculated, either numerically or
analytically, since it is easy to see that the second integral in
Eq.~(\ref{fluct2}) is subdominant and only the first integral needs to be
evaluated in order to obtain the leading behaviour of the fluctuations.

\ack

I would like to thank M. Goethe, A. Braun and M.A. Moore for many useful
discussions.

\appendix

\section{Evaluation of the interpolating Hamiltonians}
\label{approute1}

In this appendix we show the details of the derivation of the connection to
chaos. The partial derivatives in Eq.~(\ref{route1}) evaluate to
\begin{eqnarray}
\fl
  \frac{\partial}{\partial t}\log Z_t &= \frac{1}{Z_t}\Tr\left(
  \frac{\beta}{2\sqrt{t}\sqrt{N}}\sum_{i<j}J'_{ij}s_i s_j - 
  \frac{\beta}{2\sqrt{1-t}\sqrt{N}}\sum_{i<j}J_{ij}s_i s_j\right) \exp(-\beta
\mathcal H_t)\\
\fl
&= \frac{1}{2t}\sum_{i<j}J'_{ij}\frac{\partial \log Z_t}{\partial
    J'_{ij}} - \frac{1}{2(1-t)}\sum_{i<j}J_{ij}\frac{\partial \log Z_t}{\partial
    J_{ij}}
\end{eqnarray}
It remains to deal with the average over the disorder in 
\begin{eqnarray}
  \avg{\frac{\partial}{\partial t}\log Z_t\frac{\partial}{\partial
      \tau}\log Z_\tau} &= \avg{%
      \sum_{i<j,k<l}
      \left(\frac{1}{2t}J'_{ij}\frac{\partial\log Z_t}{\partial J'_{ij}} - 
            \frac{1}{2(1-t)}J_{ij}\frac{\partial\log Z_t}{\partial J_{ij}}
          \right)} \nonumber\\ 
   &\quad\times
       \left(\frac{1}{2\tau}J'_{kl}\frac{\partial\log Z_\tau}{\partial J'_{kl}}
       - \frac{1}{2(1-\tau)}J_{kl}\frac{\partial\log Z_\tau}{\partial J_{kl}}
       \right).
      \label{average1}
\end{eqnarray}

Let's look at the first term of the product under the sum,
$\avg{\frac{1}{4t\tau}J'_{ij}J'_{kl}\frac{\partial\log
      Z_t}{\partial J'_{ij}}\frac{\partial\log Z_\tau}{\partial J'_{kl}}}$.
We can integrate by parts with respect to, say, $J'_{ij}$
(a standard trick \cite{Guerra:2002}) in the form
\begin{eqnarray}
\avg{J'_{ij}J'_{kl}\bullet} &= \int \cdots dJ'_{ij}\,
e^{-{J'_{ij}}^2/2}\cdots J'_{ij}J'_{kl}\bullet \nonumber\\
&= \int \cdots dJ'_{ij}\,
e^{-{J'_{ij}}^2/2}\cdots \frac{\partial}{\partial J'_{ij}}J'_{kl} \bullet =
\avg{\frac{\partial}{\partial J'_{ij}}J'_{kl} \bullet}
\end{eqnarray} 
where the $\bullet$ stands symbollically for any function of the $J_{ij}$ and
$J'_{ij}$. The derivative can be moved to the right using the product rule so
\begin{eqnarray}
\avg{J'_{ij}J'_{kl}\bullet} &=
\avg{\left(\delta_{(ij),(kl)} + J'_{kl}\frac{\partial}{\partial
J'_{ij}}\right)}
\bullet.
\end{eqnarray}
Here, the second term can once again be treated by integration by parts, this
time with respect to $J'_{kl}$. The result is
\begin{eqnarray}
\avg{J'_{ij}J'_{kl}\bullet} &=
\avg{\left(\delta_{(ij),(kl)} +
\frac{\partial}{\partial J'_{kl}}\frac{\partial}{\partial
J'_{ij}}\right)\bullet}.
\end{eqnarray}

The same procedure can be applied to the remaining terms in
Eq.~(\ref{average1}), with the difference that the terms that mix $J$s and
$J'$s do not have the $\delta_{(ij),(kl)}$, resulting in
\begin{eqnarray}
\fl
\avg{\frac{\partial}{\partial t}\log Z_t\frac{\partial}{\partial
      \tau}\log Z_\tau} = \avg{%
      \sum_{i<j,k<l}
      \left[
      \frac{1}{4t\tau}
      \frac{\partial}{\partial J'_{kl}}
      \frac{\partial}{\partial J'_{ij}}
      \frac{\partial\log Z_t}{\partial J'_{ij}}
      \frac{\partial\log Z_\tau}{\partial J'_{kl}}
      \right.}
       - \frac{1}{4t(1-\tau)}
      \frac{\partial}{\partial J_{kl}}
      \frac{\partial}{\partial J'_{ij}}
      \frac{\partial\log Z_t}{\partial J'_{ij}}
      \frac{\partial\log Z_\tau}{\partial J_{kl}}
      \nonumber\\
       - \frac{1}{4(1-t)\tau}
      \frac{\partial}{\partial J'_{kl}}
      \frac{\partial}{\partial J_{ij}}
      \frac{\partial\log Z_t}{\partial J_{ij}}
      \frac{\partial\log Z_\tau}{\partial J'_{kl}}
       \left. + \frac{1}{4(1-t)(1-\tau)}
      \frac{\partial}{\partial J_{kl}}
      \frac{\partial}{\partial J_{ij}}
      \frac{\partial\log Z_t}{\partial J_{ij}}
      \frac{\partial\log Z_\tau}{\partial J_{kl}}
      \right]\nonumber \\
       + \frac{1}{4t\tau} \sum_{i<j}\avg{%
      \frac{\partial\log Z_t}{\partial J'_{ij}}
      \frac{\partial\log Z_\tau}{\partial J'_{ij}}}
       + \frac{1}{4(1-t)(1-\tau)} \sum_{i<j}\avg{%
      \frac{\partial\log Z_t}{\partial J_{ij}}
      \frac{\partial\log Z_\tau}{\partial J_{ij}}}
      \label{average2}
\end{eqnarray}

The derivatives which appear in this expression are related to spin averages. One
finds for example
\begin{eqnarray}
\frac{\partial\log Z_t}{\partial J_{ij}} &= \frac{\beta}{\sqrt
N}\sqrt{1-t}\langle s_i s_j\rangle_t,
\end{eqnarray}
where $\langle\cdots\rangle_t$ stands for the thermal average, to be taken with the
interpolation parameter set to $t$. Similarly, for two derivatives, one obtains
for instance
\begin{eqnarray}
\frac{\partial^2\log Z_t}{\partial J_{ij}\partial J'_{kl}} &=
\frac{\beta^2}{N}\sqrt{1-t}\sqrt{t}(\langle s_i s_j s_k s_l\rangle_t - \langle
s_i s_j\rangle_t\langle s_k s_l\rangle_t).
\end{eqnarray}
In general, each derivative with respect to a $J$ or $J'$ generates averages of
the spins with the indices involved and brings down a prefactor
$\beta\sqrt{1-t}/\sqrt{N}$ (for $J$) or $\beta\sqrt{t}\sqrt{N}$ (for $J'$). 
Fortunately, two derivatives of $\log Z_t$ are all we need because when
Eq.~(\ref{average2}) is evaluated, all terms containing higher order derivatives
drop out. This is left as an excercise for the reader. Only the following terms
survive:
\begin{eqnarray}
\fl
\avg{\frac{\partial}{\partial t}\log Z_t\frac{\partial}{\partial
      \tau}\log Z_\tau} = \frac{\beta^4}{4N^2}     
      \left(2-\frac{\sqrt{1-t}\sqrt{\tau}}{\sqrt{t}\sqrt{1-\tau}}
             -\frac{\sqrt{1-\tau}\sqrt{t}}{\sqrt{\tau}\sqrt{1-t}}\right)
             \nonumber\\
     \times        \avg{%
      \sum_{i<j,k<l}(\langle s_i s_j s_k s_l\rangle_t - \langle s_i
      s_j\rangle_t\langle s_k s_l\rangle_t)
      (\langle s_i s_j s_k s_l\rangle_\tau - \langle s_i
      s_j\rangle_\tau\langle s_k s_l\rangle_\tau)} \nonumber\\
       + \frac{\beta^2}{4N\sqrt{t\tau}} \sum_{i<j}\avg{%
      \langle s_i s_j\rangle_t\langle s_i s_j\rangle_\tau}\nonumber\\
      + \frac{\beta^2}{4N\sqrt{1-t}\sqrt{1-\tau}} \sum_{i<j}\avg{%
      \langle s_i s_j\rangle_t\langle s_i s_j\rangle_\tau}.
\end{eqnarray}
In this equation, the last term is equal to the penultimate one due to symmetry
under the exchange $t\to 1-t$ and $\tau\to 1-\tau$. In this
equation, we can let the sums run unrestrictedly over $i,j,k,l$ by introducing
a factor of $\frac 14$ for the first sum and a factor of $\frac 12$ and a
correction for the diagonal terms in the second sum, resulting in
\begin{eqnarray}
\fl
\avg{\frac{\partial}{\partial t}\log Z_t\frac{\partial}{\partial
      \tau}\log Z_\tau} = \frac{\beta^4}{16 N^2}     
      \left(2-\frac{\sqrt{1-t}\sqrt{\tau}}{\sqrt{t}\sqrt{1-\tau}}
             -\frac{\sqrt{1-\tau}\sqrt{t}}{\sqrt{\tau}\sqrt{1-t}}\right)
             \nonumber\\
     \times\avg{%
      \sum_{ijkl}(\langle s_i s_j s_k s_l\rangle_t - \langle s_i
      s_j\rangle_t\langle s_k s_l\rangle_t)
      (\langle s_i s_j s_k s_l\rangle_\tau - \langle s_i
      s_j\rangle_\tau\langle s_k s_l\rangle_\tau)} \nonumber\\
       + \frac{\beta^2}{4N\sqrt{t\tau}} \left(\sum_{ij}\avg{%
      \langle s_i s_j\rangle_t\langle s_i s_j\rangle_\tau} - N\right).
      \label{average3}
\end{eqnarray}
We can write $\sum_{ij}\langle s_i s_j\rangle_t\langle s_i s_j\rangle_\tau$ as
$N^2\langle \left(\frac 1N\sum_i s_i^{1t} s_i^{3\tau}\right)^2\rangle$,  i.e.\
as a square of the spin overlap 
\begin{eqnarray}
q_{13}(t,\tau) &= \frac 1N \sum_i s_i^{1t} s_i^{3\tau}
\end{eqnarray} 
between two replicas labelled $1$ and $3$ \textit{with different interpolation
parameters} $t$ and $\tau$. The labels $1$ and $3$ have been chosen because
we will shortly need two more replicas which will be assigned the labels
$2$ and $4$. Replicas $1$ and $2$ are then understood to have interpolation
parameter $t$ and replicas $3$ and $4$ to have parameter $\tau$. The angular
brackets $\langle\cdots\rangle$ without subscript indicate the thermal 
average of a system comprising independent replicas with interpolation 
parameters $t$ and $\tau$.

A similar decomposition in replicas can be made in the first part of
Eq.~(\ref{average3}), but here two more replicas are needed. We find
\begin{eqnarray}
\fl
\frac{1}{N^4}\sum_{ijkl}(\langle s_i s_j s_k s_l\rangle_t - \langle s_i
      s_j\rangle_t\langle s_k s_l\rangle_t)
      (\langle s_i s_j s_k s_l\rangle_\tau - \langle s_i
      s_j\rangle_\tau\langle s_k s_l\rangle_\tau) \nonumber \\
      =
\frac{1}{N^4}\sum_{ijkl}(\langle s_i^{1,t} s_j^{1,t} s_k^{1,t}
s_l^{1,t}\rangle-\langle s_i^{1,t} s_j^{1,t}\rangle \langle s_k^{2,t}
s_l^{2,t}\rangle)(\langle s_i^{3,\tau} s_j^{3,\tau} s_k^{3,\tau}
s_l^{3,\tau}\rangle-\langle s_i^{4,\tau} s_j^{4,\tau}\rangle \langle
s_k^{3,\tau} s_l^{3,\tau}\rangle) \nonumber\\
=\langle(q_{13}^2 - q_{14}^2)(q_{13}^2-q_{23}^2)\rangle \nonumber\\
=\frac{1}{N^4}\left\langle\left(
\sum_{ij}(s_i^{1,t} s_j^{1,t}-\langle s_i^{1,t}s_j^{1,t}\rangle)
(s_i^{3,\tau}s_j^{3,\tau}-\langle s_i^{3,\tau}s_j^{3,\tau}\rangle)\right)^2 
\right\rangle \ge 0.
\end{eqnarray}
As a by-product, we see in the last line that the expression $\langle(q_{13}^2 -
q_{14}^2)(q_{13}^2-q_{23}^2)\rangle$ is nonnegative.
The end result is finally
\begin{eqnarray}
\fl
\avg{\frac{\partial}{\partial t}\log Z_t\frac{\partial}{\partial
      \tau}\log Z_\tau} = \frac{N^2\beta^4}{16}     
      \left(2-\frac{\sqrt{1-t}\sqrt{\tau}}{\sqrt{t}\sqrt{1-\tau}}
             -\frac{\sqrt{1-\tau}\sqrt{t}}{\sqrt{\tau}\sqrt{1-t}}\right)
        \avg{\langle(q_{13}^2-q_{14}^2)(q_{13}^2-q_{23}^2)\rangle}\nonumber\\
      + \frac{N \beta^2}{4\sqrt{t\tau}} \left(\avg{%
     \langle q_{13}^2\rangle} - \frac 1N\right).
     \label{average4}
\end{eqnarray}

\section*{References}

\bibliographystyle{unsrt}
\bibliography{LiteraturDB,cond-mat}

\begin{thebibliography}{10}

\bibitem{Biroli:2007}
G.~Biroli, J.-P. Bouchaud, and M.~Potters.
\newblock Extreme value problems in random matrix theory and other disordered
  systems.
\newblock {\em J. Stat. Mech.}, 2007:P07019.

\bibitem{Sherrington:1975}
D.~Sherrington and S.~Kirkpatrick.
\newblock Solvable model of a spin-glass.
\newblock {\em Phys. Rev. Lett.}, 35:1792, 1975.

\bibitem{Cabasino:1988}
S~Cabasino, E~Marinari, P~Paolucci, and G~Parisi.
\newblock Eigenstates and limit cycles in the {SK} model.
\newblock {\em J. Phys. A}, 21(22):4201--4210, 1988.

\bibitem{Bouchaud:2003}
J.-P. Bouchaud, F.~Krzakala, and O.~C. Martin.
\newblock Energy exponents and corrections to scaling in {I}sing spin glasses.
\newblock {\em Phys. Rev. B}, 68:224404, 2003.

\bibitem{Palassini:2003}
M.~Palassini.
\newblock Ground-state energy fluctuations in the {S}herrington-{K}irkpatrick
  model.
\newblock cond-mat/0307713, 2003.

\bibitem{Andreanov:2004}
A.~Andreanov, F.~Barbieri, and O.~C. Martin.
\newblock Large deviations in spin-glass ground-state energies.
\newblock {\em Eur. Phys. J. B}, 41(3):365--375, 2004.

\bibitem{Boettcher:2005a}
S.~Boettcher.
\newblock Extremal optimization for {S}herrington-{K}irkpatrick spin glasses.
\newblock {\em Eur. Phys. J. B}, 46:501--505, 2005.

\bibitem{Katzgraber:2005}
H.~G. Katzgraber, M.~K{\"o}rner, F.~Liers, M.~J{\"u}nger, and A.~K. Hartmann.
\newblock Universality-class dependence of energy distributions in spin
  glasses.
\newblock {\em Phys. Rev. B}, 72:094421, 2005.

\bibitem{Pal:2006}
K.~F. P\'al.
\newblock Hysteretic optimization for the {S}herrington-{K}irkpatrick spin
  glass.
\newblock {\em Physica A}, 367:261--268, 2006.

\bibitem{Aspelmeier:2003a}
T.~Aspelmeier, M.~A. Moore, and A.~P. Young.
\newblock Interface energies in {I}sing spin glasses.
\newblock {\em Phys. Rev. Lett.}, 90(12):127202, 2003.

\bibitem{Crisanti:1992}
A.~Crisanti, G.~Paladin, H.-J. Sommers, and A.~Vulpiani.
\newblock Replica trick and fluctuations in disordered systems.
\newblock {\em J. Phys. I France}, 2:1325--1332, 1992.

\bibitem{Aspelmeier:2007}
T.~Aspelmeier, A.~Billoire, E.~Marinari, and M.~A. Moore.
\newblock Finite size corrections in the {S}herrington-{K}irkpatrick model.
\newblock arXiv:0711.3445v1 [cond-mat.dis-nn], submitted to J. Phys. A, 2007.

\bibitem{Wehr:1990}
J.~Wehr and M.~Aizenman.
\newblock Fluctuations of extensive functions of quenched random couplings.
\newblock {\em J. Stat. Phys.}, 60(3-4):287, 1990.

\bibitem{Aspelmeier:2003b}
T.~Aspelmeier and M.~A. Moore.
\newblock Free energy fluctuations in {I}sing spin glasses.
\newblock {\em Phys. Rev. Lett.}, 90(17):177201, 2003.

\bibitem{Aspelmeier:2007a}
T.~Aspelmeier.
\newblock Free energy fluctuations and chaos in the {S}herrington-{K}irkpatrick
  model.
\newblock arXiv:0712.3586v1 [cond-mat.dis-nn], accepted by Phys. Rev. Lett.,
  2007.

\bibitem{Aspelmeier:2008bpre}
T.~Aspelmeier.
\newblock Bond chaos in the {S}herrington-{K}irkpatrick model.
\newblock arXiv:????.????, see this mailing, 2008.

\bibitem{Mezard:1987}
M.~M\'ezard, G.~Parisi, and M.A. Virasoro.
\newblock {\em Spin Glass Theory and Beyond}.
\newblock World Scientific, Singapore, 1987.

\bibitem{Parisi:1993}
G.~Parisi, F.~Ritort, and F.~Slanina.
\newblock Critical finite-size corrections for the {S}herrington-{K}irkpatrick
  spin glass.
\newblock {\em J. Phys. A}, 26:247--259, 1993.

\bibitem{Yeo:2005}
J.~Yeo, M.~A. Moore, and T.~Aspelmeier.
\newblock Nature of perturbation theory in spin glasses.
\newblock {\em J. Phys. A}, 38(18):4027--4045, 2005.

\bibitem{Billoire:2006}
A.~Billoire.
\newblock Numerical estimate of the finite-size corrections to the free energy
  of the {S}herrington-{K}irkpatrick model using {G}uerra-{T}oninelli
  interpolation.
\newblock {\em Phys. Rev. B}, 73:132201, 2006.

\bibitem{Guerra:2002}
F.~Guerra and F.~L. Toninelli.
\newblock The thermodynamic limit in mean field spin glass models.
\newblock {\em Commun. Math. Phys.}, 230:71--79, 2002.

\bibitem{Parisi:2000}
G.~Parisi and F.~Ricci-Tersenghi.
\newblock On the origin of ultrametricity.
\newblock {\em J. Phys. A}, 33(1):113--129, 2000.

\bibitem{Guerra:1996}
F.~Guerra.
\newblock About the overlap distribution in mean field spin glass models.
\newblock {\em Int. J. Mod. Phys. B}, 10:1675, 1996.

\end{thebibliography}

\end{document}